\documentclass{webofc}

\usepackage[varg]{txfonts}   
\usepackage{hyperref}
\usepackage{url}
\hypersetup{colorlinks=true,citecolor=blue,urlcolor=blue,linkcolor=blue}
\begin{document}
\title{Visualizing How Jet Structure Shapes Jet Wakes}
%
%

\author{\firstname{Arjun} \lastname{Kudinoor}\inst{1,2}\fnsep\thanks{Speaker at HP24; \email{kudinoor@mit.edu}} \and
        \firstname{Daniel} \lastname{Pablos}\inst{3}\fnsep\thanks{\email{daniel.pablos@usc.es}} \and
        \firstname{Krishna} \lastname{Rajagopal}\inst{1}\fnsep\thanks{\email{krishna@mit.edu}}
}

\institute{Center for Theoretical Physics, Massachusetts Institute of Technology, Cambridge, MA 02139
\and
DAMTP, University of Cambridge, Cambridge, CB3 0WA, UK 
\and
IGFAE, Universidade de Santiago de Compostela, E-15782 Galicia-Spain
          }

\abstract{
The ATLAS Collaboration has developed a method to analyze large-radius jets composed of skinny $R=0.2$ subjets in heavy-ion collisions. 
We first 
demonstrate that the 
measurements pioneered by ATLAS constrain the value of $L_{\rm res}$, the resolution length of QGP --- and rule out any picture in which an entire parton shower loses energy coherently as 
a single entity. We then analyze the response of the medium to the passage of large-radius $R=2$ jets containing two skinny subjets in gamma-jet events. We introduce novel jet-shape observables that allow us to visualize 
how the internal structure of large-radius jets shapes the wakes they excite in the QGP.
We find that even when the subjets are $\sim 0.8$ radians apart, 
the angular shape of the soft hadrons originating from their wake forms
a single broad structure. Only when the two subjets are even farther apart are 
two sub-wakes revealed. We show that the way in which jet structure shapes the structure of jet-induced wakes can be visualized with similar clarity in experiments by using only low-$p_T$ hadrons.
The observables we introduce offer a new and distinctive way of seeing jet-induced wakes -- and wake substructure -- in heavy-ion collision data.
}
\maketitle
%

\section{The Hybrid Strong/Weak Coupling Model} \label{sec:hybrid-model}
The hybrid strong/weak coupling model, or simply the Hybrid Model, is a theoretical framework for jet quenching designed to describe the multi-scale processes of jet production and evolution through a strongly coupled plasma. The Hybrid Model treats the weakly coupled physics of jet production and hard jet evolution perturbatively. Parton splittings that result in a jet shower are determined by the high-$Q^2$, perturbative, DGLAP evolution equations, implemented using PYTHIA 8. The non-perturbative interactions between partons in a jet shower and the droplet of QGP through which they propagate dictate that these partons lose energy to the strongly coupled plasma. In the Hybrid Model, each parton in a jet shower loses energy to the plasma as determined by an energy loss formula, calculated holographically in Refs.~\cite{Chesler:2014jva,Chesler:2015nqz} and detailed in Ref.~\cite{Kudinoor:2025ilx}.


Since energy and momentum must be conserved, the momentum and energy lost by a parton is deposited into the plasma, exciting a hydrodynamic wake in the expanding, flowing, and cooling droplet of the liquid QGP. One can think of this wake as a portion of the medium that is pulled in the direction of the energetic partons. When the QGP droplet and the wake(s) within it reach the freezeout hypersurface, they hadronize into thousands of soft hadrons, a subset of which are the result of the wake(s) hadronizing at freezeout. In the Hybrid Model, the spectrum of hadrons belonging to a jet-induced wake is determined by employing the Cooper--Frye prescription to the jet-induced perturbation of the stress-energy tensor of the strongly coupled liquid QGP, assuming that the background fluid is longitudinally boost invariant and that the jet-induced perturbation stays close in rapidity to the rapidity of the jet \cite{Casalderrey-Solana:2016jvj}. The rationality of these assumptions are discussed in Section 2.1 of Ref.~\cite{Kudinoor:2025ilx}. For a jet with rapidity $y_j$ and azimuthal angle $\phi_j$ that has lost transverse momentum $\Delta p_T$ and energy $\Delta E$ to the plasma produces a wake with the spectrum given by
\begin{equation}
\label{eq:onebody}
\begin{split}
E\frac{\rm d\Delta N}{\rm d^3p}=&\frac{1}{32 \pi} \, \frac{m_T}{T^5} \, \textrm{cosh}(y-y_j)  e^{-\frac{m_T}{T}\, \textrm{cosh}(y-y_j)} \\
 &\times \Bigg\{ p_{T} \Delta p_{T} \cos (\phi-\phi_j) +\frac{1}{3}m_T \, \frac{\Delta E}{\textrm{cosh}(y_j)} \, \textrm{cosh}(y-y_j) \Bigg\} \, ,
\end{split}
\end{equation}
where $y$, $\phi$, $p_T$, and $m_T$ are the rapidity, azimuthal angle, transverse momentum, and transverse mass of the generated wake-hadrons.

Notice that the above spectrum can be negative for some ranges of rapidity and azimuthal angle. One might ask what this means. Since a jet pulls some amount of QGP in its direction of propagation, if you compare the freezeout of a droplet of QGP containing a jet-induced wake to a QGP droplet without a wake, you will observe an excess of soft particles in the direction of the jet's propagation and a depletion of soft particles in the direction opposite the jet's propagation. In momentum space, this manifests itself as soft particles with positive momentum in the direction of the jet and soft particles with negative momentum in the direction opposite the jet. We call the collections of these particles the positive wake and negative wake, respectively. The negative wake resulting from the depletion of soft particles in the direction opposite the jet results in the negative regimes of the above spectrum of wake-hadrons.

In addition to jet-induced wakes, there are many other physical effects that can affect the response of the medium to energetic parton showers that traverse it. One such effect is the \textit{QGP resolution length} $L_{\rm res}$, which is the length scale below which the medium cannot resolve two partons within a jet shower as different sources of energy loss. So, two partons within the same jet shower lose energy independently to and deposit energy independently into the medium if and only if they are separated by a length larger than $L_{\rm res}$. If they are separated by a length less than $L_{\rm res}$, they lose energy to and deposit energy into the medium as if they were a single object. The implementation of a QGP resolution length within the Hybrid Model is detailed in Ref.~\cite{Kudinoor:2025ilx}. We restrict to studying three values of $L_{\rm res}$ --- namely 0, $2/(\pi T)$, and $\infty$. $L_{\rm res} = 0$ corresponds to fully incoherent energy loss, with every parton in the shower resolved, while $L_{\rm res} = \infty$ corresponds to fully coherent energy loss, with the entire shower losing energy as if it were a single object.

\begin{figure}
\centering
\includegraphics[width=13.5cm,clip]{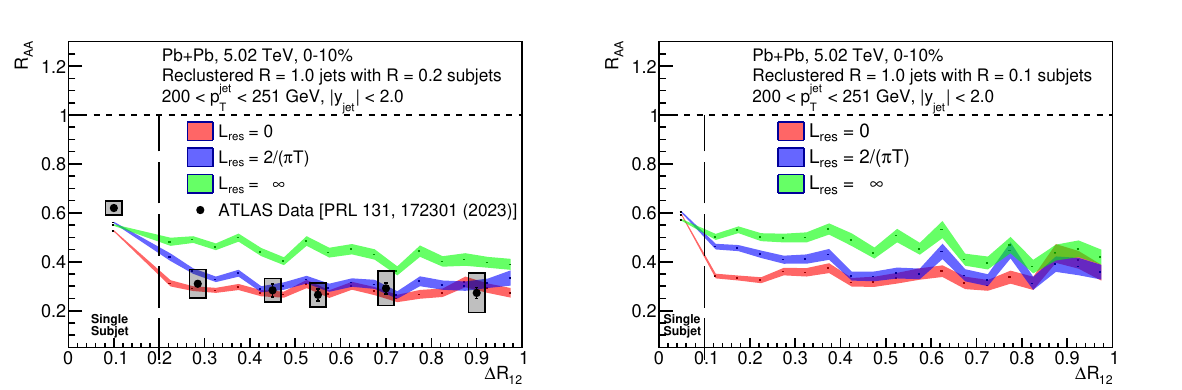}
\caption{$R_{\rm AA}$ as a function of $\Delta R_{12}$ for large-radius $R = 1.0$ jets with $200 < p_T < 251$ GeV, reconstructed using $R = 0.2$ subjets as constituents (left) and using $R = 0.1$ subjets as constituents (right). The leftmost bin corresponds to $R = 1.0$ jets with a single subjet.  The colored bands show the results of
Hybrid Model calculations with $L_{\rm res} = 0$ (red), $2/(\pi T)$ (blue) and $\infty$ (green). ATLAS
experimental measurements from Ref.~\cite{ATLAS:2023hso} are depicted using point markers, upon which the vertical bars indicate statistical uncertainties and the shaded
boxes indicate systematic uncertainties extracted from Ref.~\cite{ATLAS:2023hso}.}
\label{fig:deltar12}
\end{figure}

\section{QGP Resolution Length Effects on Large-Radius Jet Suppression}

\label{sec:lres}
In Ref.~\cite{ATLAS:2023hso}, the ATLAS Collaboration pioneered a method to identify and analyze large-radius jets reconstructed from skinny subjets. Following their method, we first reconstruct anti-$k_t$ $R = 0.2$ jets with $|\eta| < 3.0$ and $p_T > 35$ GeV in inclusive jet events. We refer to these $R = 0.2$ jets as ``skinny subjets", which we then use as the constituents for reconstructing anti-$k_t$ $R = 1.0$ jets with $|y| < 2.0$ and $p_T > 158$ GeV. Finally, we recluster the $R = 1.0$ jets using the $k_t$-recombination algorithm to obtain the observable $\Delta R_{12} \equiv \sqrt{\Delta y_{12}^2 + \Delta \phi_{12}^2}$, defined as the angular separation between the two skinny subjet constituents involved in the final reclustering step of the $R = 1.0$ jet. Since the $k_t$ algorithm tends to combine the hardest constituents of a jet last, $\Delta R_{12}$ corresponds to the angular scale of the hardest splitting between the anti-$k_t$ subjets within a large-radius jet. For a large-radius jet composed of a single skinny subjet, $\Delta R_{12} \equiv 0$.

We calculated $R_{\rm AA}$ for large-radius $R = 1.0$ jets as a function of their $p_T$ and $\Delta R_{12}$, for three different values of $L_{\rm res}$. Ref.~\cite{Kudinoor:2025ilx} includes an analysis of large-radius jet suppression as a function of jet-$p_T$. In particular, we find that $R = 1.0$ jets composed of a single skinny subjet are less suppressed than $R = 1.0$ jets with multiple subjets for all values of $L_{\rm res}$. Since a QGP droplet with $L_{\rm res} < \infty$ can resolve partons within a jet shower, jets containing multiple subjets will contain multiple sources of energy loss and therefore will be suppressed more than jets containing only a single subjet. However, the story is not so obvious when $L_{\rm res} = \infty$ because such a plasma will never resolve multiple sources of energy loss within a jet shower.
Appendix A of Ref.~\cite{Kudinoor:2025ilx} details how initial state radiation plays contributes to the additional suppression experienced by $R = 1.0$ jets with multiple subjets.

The left subfigure of Fig.~\ref{fig:deltar12} shows $R_{\rm AA}$ as a function of $\Delta R_{12}$ for $R = 1.0$ jets with $200 < p_T < 251$ GeV reconstructed from $R = 0.2$ skinny subjets. The colored bands show Hybrid Model calculations for $L_{\rm res} = 0$ (red), $2/(\pi T)$ (blue), and $\infty$ (green). Our Hybrid Model calculations demonstrate good agreement with ATLAS' measurements for $L_{\rm res} = 0$ and $2/(\pi T)$. However, ATLAS' measurements disfavor an infinite QGP resolution length, meaning they disfavor fully coherent energy loss. We also note that $R_{\rm AA}$ for $R = 1.0$ jets with multiple subjets ($\Delta R_{12} > 0$) is lower than $R_{\rm AA}$ for $R = 1.0$ jets with a single skinny subjet by a factor of $\sim 2$. This factor of 2 is due to large-radius jet sample being dominated by $R = 1.0$ jets containing two skinny subjets \cite{Kudinoor:2025ilx}, i.e. two sources of energy loss. Furthermore, we note that $R_{\rm AA}$ is constant as a function of $\Delta R_{12} > 0$ when $L_{\rm res} = 0$ because the medium is able to resolve all skinny subjets within a large-radius jet as independent sources of energy loss, regardless of how far apart they are in angle. However, when $L_{\rm res} = 2/(\pi T)$, $R_{\rm AA}$ is not flat when $0.2 < \Delta R_{12} < 0.4$. This is because for smaller $\Delta R_{12}$, it is difficult for the QGP to resolve the hard partons from which the skinny subjets originate since partons in the jet shower that are separated by a distance less than $L_{\rm res}$ will lose energy to the plasma as if they were one unresolved parton.

It would be interesting to extend this measurement down to values of $\Delta R_{12} \approx 0.1$ by using skinnier subjets with $R = 0.1$. The right subfigure of Fig.~\ref{fig:deltar12} shows that using a subjet radius of $R = 0.1$ allows one to access a larger range of low-$\Delta R_{12}$ with which to discriminate between $L_{\rm res} = 0$ and $L_{\rm res} = 2/(\pi T)$. Therefore, accessing lower values of $\Delta R_{12}$ in experiment would gives us the ability to further constrain the value of the QGP resolution length.

\section{Imaging the Structure of Large-Radius Jet Energy Loss} \label{sec:wake-shape}
\begin{figure}
\centering
\sidecaption
\includegraphics[width=4cm,clip]{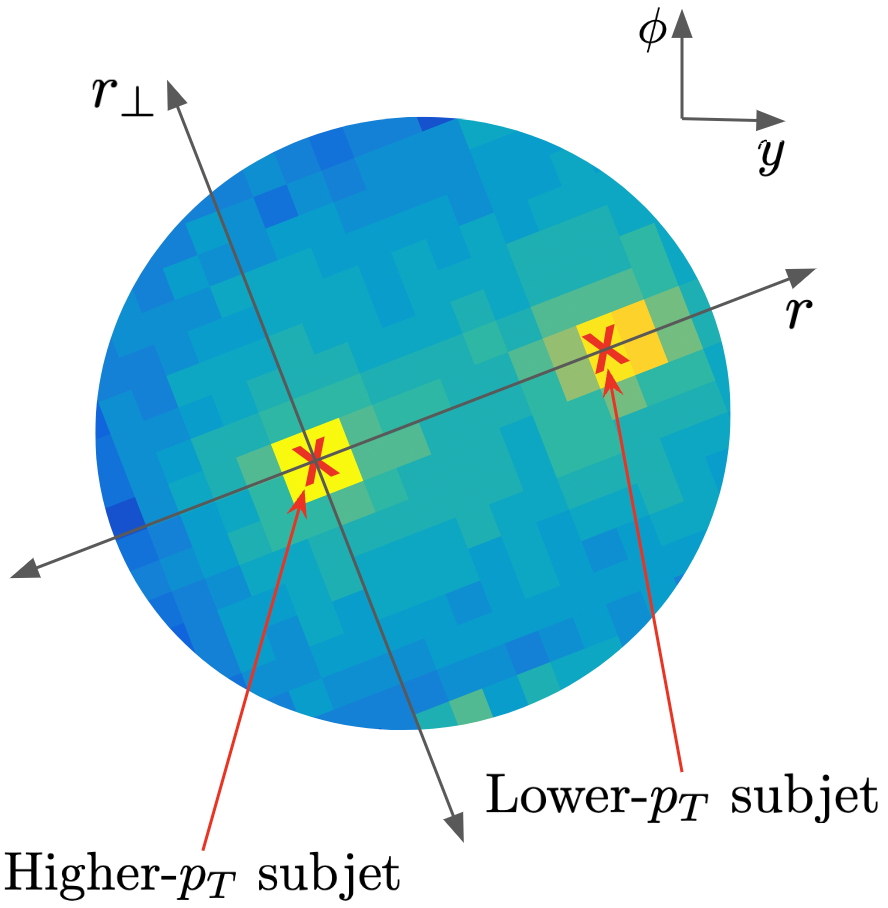}
\caption{A schematic diagram of the $(r, r_\perp)$ coordinate system for a given large-radius jet with two skinny subjets (the red X's). The origin of the $(r,r_\perp)$ coordinates is centered on the higher-$p_T$ subjet, with the $r$-axis pointing through the center of the lower-$p_T$ subjet. The $r_\perp$ axis is perpendicular to the $r$-axis.}
\label{fig:setup}
\end{figure}

In the previous Section, we saw that in a droplet of QGP with $L_{\rm res} = 0$, the fraction of energy that partons in a jet shower lose to the plasma is independent of the separations between those partons. So, the suppression experienced by large-radius jets with multiple skinny subjets is constant as a function of the separation between the subjets when $L_{\rm res} = 0$. However, this does not mean that the structural redistribution of the energy and momentum lost by each large-radius jet is independent of the angular separation between its subjets. In this section, our goal is to visualize how the substructure of a large-radius jet shapes the substructure of the wake it excites in the plasma. In Ref.~\cite{Kudinoor:2025ilx} and as detailed below, we introduce novel jet shape observables to achieve this goal.

For a large-radius jet of radius $R_{\rm large}$ with two skinny subjets, we first construct a new coordinate system $(r, r_\perp)$ whose origin is fixed on the axis of the higher-$p_T$ subjet. We then define the $r$-axis to point through the center of the lower-$p_T$ subjet, and the $r_\perp$-axis to be perpendicular to the $r$-axis. Fig.~\ref{fig:setup} shows a schematic diagram of this coordinate system. The $(r, r_\perp)$ coordinate system is constructed so that the axes of the two subjets lie on the $r$-axis. This enables us to study the region between and around these two subjets. To do so, we define a new differential jet shape observable at location $(r, r_\perp)$ as the fraction of a large-radius jet's hadronic energy contained within a $\delta r \times \delta r_\perp$ box centered at $(r, r_\perp)$:

\begin{equation} \label{eq:newjetshape}
    \rho(r, r_\perp) \equiv \frac{1}{N_{\rm jets}} \frac{1}{\delta r \text{ } \delta r_\perp} \sum_{\rm jets} \frac{\sum_{i \in (r \pm \delta r /2, r_\perp \pm \delta r_\perp / 2)} p_T^i}{p_T^{\rm jet}},
\end{equation}
where $p_T^{\rm jet}$ is the $p_T$ of the large-radius jet, and $i$ runs over \textit{all} the hadrons within a radius of $\Delta R \equiv \sqrt{\Delta y^2 + \Delta \phi^2} = R_{\rm large}$ around the axis of the large-radius jet (and not only the hadrons within the two skinny subjets). Eq.~\ref{eq:newjetshape} is normalized by the number of large-radius jets with two skinny subjets that pass the selection criteria (detailed below), which will include a condition on the separation $\Delta y_{12}$ in rapidity between the two skinny subjets.

\begin{figure}
\centering
\includegraphics[width=13cm,clip]{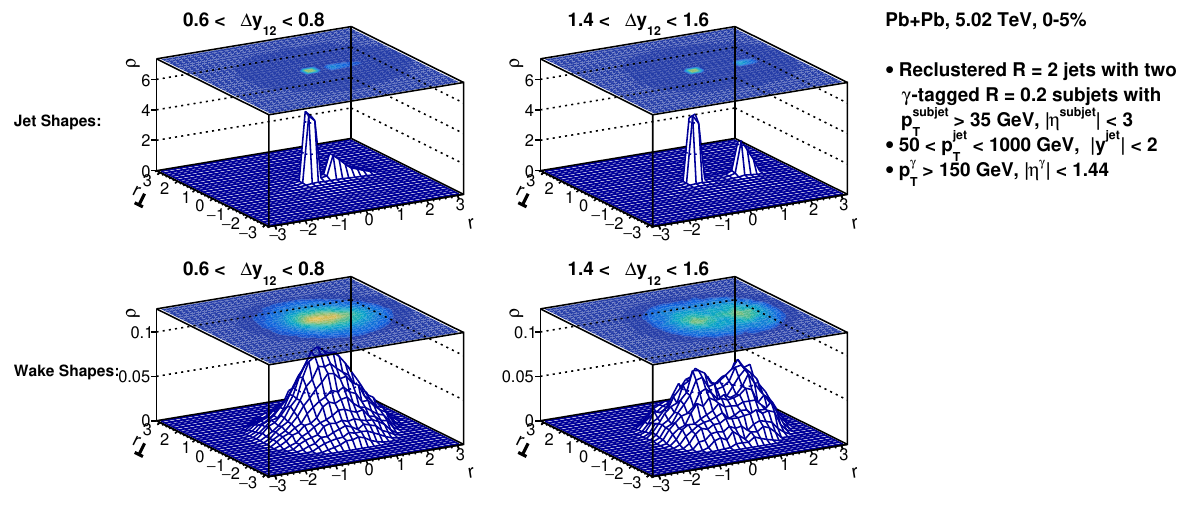}
\caption{Shapes of reclustered $R = 2.0$ jets (top row) and their wakes (bottom row) with two anti-$k_t$ $R=0.2$ subjets in $\gamma$-tagged events, calculated using hadrons within a radius of 2.0 around the $R = 2.0$ jet-axis. The two columns show different ranges of the separation $\Delta y_{12}$ in rapidity between the two skinny subjets --- namely $0.6 < \Delta y_{12} < 0.8$ (left column) and $1.4 < \Delta y_{12} < 1.6$ (right column).}
\label{fig:2dshapes}
\end{figure}

We study our novel jet shape observables in the context of large-radius $R =2$ jets with $R = 0.2$ skinny subjets in $\gamma$-tagged events, reconstructed in the same way as explained in Sec.~\ref{sec:lres}. The very large jet-radius of $R = 2$ gives us access to a large phase space to study how the internal structure of large-radius jets
shape their wakes. We restrict to $\gamma$-tagged events because the jet shapes of large-radius jets selected in inclusive jet events may be contaminated by the wake(s) of jets in the hemisphere opposite to the ones we select. See Appendix B of Ref.~\cite{Kudinoor:2025ilx} for more details on this and how the internal structure of large-radius jets shape their wakes in inclusive jet events. In our analysis, we require that the $R = 0.2$ subjets satisfy $p_T > 35$ GeV, $|\eta| < 3.0$, and are $\Delta \phi_{{\rm jet}, \gamma} > 2\pi/3$ away from a high-energy photon with $p_T > 150$ GeV and $|\eta^{\gamma}| < 1.44$ that is ``isolated", which we define as having less than 5 GeV of transverse energy in a cone of $R = 0.4$ around the photon. Furthermore, we select only those $R = 2$ jets that have two $\gamma$-tagged subjets and which satisfy $p_T > 50$ GeV and $|y| < 2.0$.

Note that if we let $i$ run over the subset of hadrons that are produced by the wakes of energetic partons in the jet shower, sampled from the spectrum given by Eq.~\ref{eq:onebody}, then $\rho(r, r_\perp)$ enables us to visualize the wakes excited by large-radius jets with two subjets. Fig.~\ref{fig:2dshapes} shows the shapes of $R = 2$ jets with two $\gamma$-tagged subjets (upper panels) and their wakes (bottom panels) for two ranges of separation $\Delta y_{12}$ in rapidity between the subjets. In the upper panels, the jet shapes show distinct, tall, and sharp peaks at the locations of the two subjets. The jet shapes are dominated by the hadrons coming from the fragmentation of the collinear parton showers. However, the wakes produced by these large-radius jets tell a different story. Even when two subjets are $\sim 0.6 - 0.8$ radians apart in rapidity, they produce a single, soft, broad wake (as can be seen in the bottom left panel of Fig.~\ref{fig:2dshapes}. Recall that in this section, our jets traverse a plasma with $L_{\rm res} = 0$. So, each subjet sources its own wake. However, by the time hadrons emerge from these wakes at freezeout, their angular distribution is so wide that when the wakes superpose in the final hadronic state, the hadrons from two independently quenched, well-separated subjets can form a single broad structure. In fact, in Ref.~\cite{Kudinoor:2025ilx} we show that as long as the two subjets are separated by $\Delta y_{12} \lesssim 1.2$, a single, soft, broad wake is produced. Only when the subjets are separated in rapidity by at least 1.2 do two sub-wakes emerge. For example, the bottom right panel of Fig.~\ref{fig:2dshapes} shows that the wake produced by an $R = 2$ jet with two $\gamma$-tagged subjets separated by $1.4 < \Delta y_{12} < 1.6$ contains two sub-wakes.

\begin{figure}
\centering
\includegraphics[width=13cm,clip]{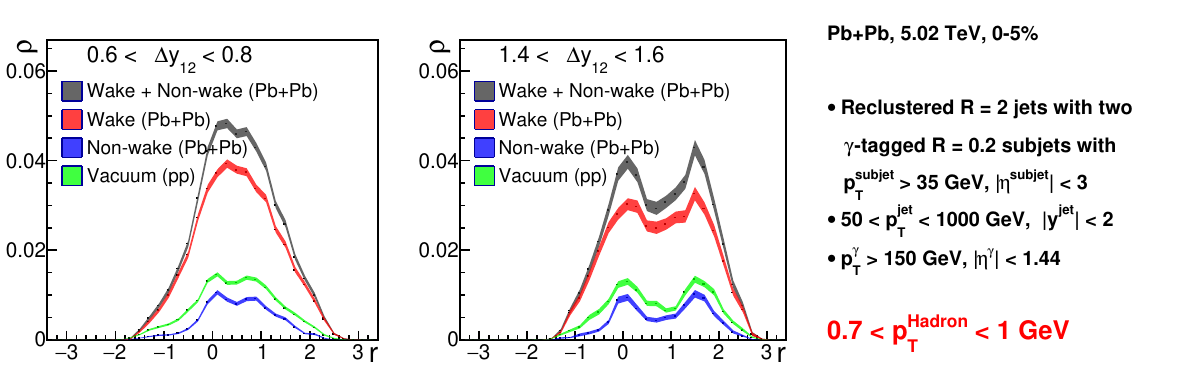}
\caption{Projections of the 2-dimensional jet shapes of reclustered $R = 2.0$ jets with two skinny $\gamma$-tagged subjets onto the $r$-axis, calculated using only those hadrons with $0.7 < p_T < 1.0$ GeV within a radius of $\Delta R = 2.0$ around the reclustered $R = 2.0$ jet's axis. The bands correspond to jet shape projections calculated using only jet wake particles in PbPb collisions (red), non-wake particles from jet fragmentation in PbPb collisions (blue), all particles from jets in PbPb collisions (gray), and particles from jets in vacuum (green). The two panels show different ranges of the separation $\Delta y_{12}$ in rapidity between the two skinny subjets --- namely $0.6 < \Delta y_{12} < 0.8$ (left) and $1.4 < \Delta y_{12} < 1.6$ (right).}
\label{fig:1dshapes}
\end{figure}

In experimental data, hadrons are not labeled as originating from the freeze-out of the wake or originating from the fragmentation of a parton shower. However, if we restrict our jet shape observable in Eq.~\ref{eq:newjetshape} to use only hadrons with low transverse momentum around the large-radius jet's axis, we can maximize the relative contribution of hadrons originating from the wake relative to hadrons originating from parton showers. Fig.~\ref{fig:1dshapes} shows projections of the 2-dimensional jet shapes of large-radius $R = 2.0$ jets with two $R = 0.2$ $\gamma$-tagged subjets onto the $r$-axis, calculated using only those hadrons with $0.7 < p_T < 1$ GeV. We see that the overall jet shape (gray) of these low-$p_T$ hadrons is dominated by the contributions coming from hadrons originating from the freezeout of large-radius jet wakes (red). Furthermore, the features of large-radius wake substructure apparent in Fig.~\ref{fig:2dshapes} and discussed above are still visible in the 1-dimensional projections of the experimentally measurable jet shapes (gray) calculated using only the low-$p_T$ hadrons. Therefore, our novel jet shape observables and kinematic restrictions on the hadrons used yield promising avenues for experimentalists to visualize how the internal structure of large-radius jets shapes the structure of their wakes.\\


Research supported by the U.S. Department of Energy, Office of Science, Office of Nuclear Physics under grant Contract Number DE-SC0011090.
DP is funded by the European Union's Horizon 2020 research and innovation program under the Marie Sk\l odowska-Curie grant agreement No 101155036 (AntScat), by the European Research Council project ERC-2018-ADG-835105 YoctoLHC, by the Spanish Research State Agency under project 
PID2020-119632GB-I00, by Xunta de Galicia (CIGUS Network of Research Centres) and the European Union, and by Unidad de Excelencia Mar\'ia de Maetzu under project CEX2023-001318-M.
ASK is supported by a National Science Foundation Graduate Research Fellowship Program under Grant No. 2141064.




%
\bibliography{bibliography} 
%
%

\end{document}